\documentclass[a4paper,fleqn,useAMS,usenatbib]{mnras}
  \usepackage{amsmath}
  \usepackage{amssymb}
  \usepackage{mathptmx}
  \usepackage{txfonts}
  \usepackage{graphicx}
  \usepackage[T1]{fontenc}
  \usepackage{ae,aecompl}

  \title[Intra-day variability of S5 0716+714]{Simultaneous optical monitoring of BL Lacertae object S5 0716+714 with high temporal resolution}
\author[Man et al.]
{Zhongyi Man$^{1,2,3}$, Xiaoyuan Zhang$^{1}$, Jianghua Wu$^{1}$
\thanks{E-mail:jhwu@bnu.edu.cn} \thanks{Corresponding author} and Qirong
Yuan$^{4}$\\
$^{1}$Department of Astronomy, Beijing Normal University, Beijing 100875, China\\$^{2}$Department of Astronomy, Physics School, Peking University, Beijing 100871, China\\
$^{3}$Kavli Institute for Astronomy and Astrophysics (KIAA), Peking
University, Beijing 100871, China\\
$^{4}$Department of Physics and Institute of Theoretical Physics, Nanjing
Normal University, Nanjing 210046, China}

\begin{document}

\date{in original form 2015 August 3}

%\pagerange{\pageref{firstpage}--\pageref{lastpage}} \pubyear{2015}

\maketitle

\label{firstpage}

\begin{abstract}
We have monitored the BL Lacertae object S5 0716+714 simultaneously in the
{\it B}, {\it R} and {\it I} bands on three nights in November 2014. The
average time resolution is quite high (73s, 34s, 58s for the filters {\it B},
{\it R} and {\it I}), which can help us trace the profile of the variation and
search for the short inter-band time delay. Intra-day variability was about
0.1 mag on the first two nights and more than 0.3 mag on the third. A
bluer-when-brighter color behavior was found. An clear loop path can be seen
on the color-magnitude diagram of the third night, revealing possible time
delays between variations at high and low energies. It is the first time that
the intra-day spectral hysteresis loop has been found so obviously in the
optical band. We used the interpolated cross-correlation function method to
further confirm the time delay and calculated the values of lag between light
curves at different wavelengths on each night. 
On the third night, variations in the {\it R} and {\it B} bands is
approximately 1.5 minutes lagging behind the {\it I} band. Such optical time
delay is probably due to the interplay of different processes of electrons in
the jet of the blazar.
\end{abstract}

\begin{keywords}
BL Lacertae objects : individual (S5 0716+714) -- galaxies : active -- galaxies : photometry
\end{keywords}

%%%%%%%%%%%%%%%%      content       %%%%%%%%%%%%%

\section{Introduction}
As the most extreme class of active galactic nucleis (AGN), blazars exhibit
rapid and large amplitude variability from radio to gamma-ray bands (e.g.
\citealt{2003A&A...402..151R}; \citealt{2008A&A...481L..79V};
\citealt{2012ApJ...756...13B}; \citealt{2014A&A...562A..79S}), high and
variable polarization (e.g., \citealt{2008PASJ...60L..37S};
\citealt{2014ApJ...781L...4G}) and a non-thermal continuum. In general, a
blazar is believed to be powered by a central black hole with its accretion
disk producing and accelerating relativistic jets outwards. This may explain
its extreme behavior considering the jet is oriented very close to the line of
sight (\citealt{1995PASP..107..803U}). A typical spectral energy distribution (SED) of BL Lacs is constituted of two components, according to the peak frequency of the low-energy component, we can classify BL Lacs as low/intermediate/high-frequency-peaked BL Lac objects (hereafter, LBL/IBL/HBL). 

\begin{figure}
\includegraphics[width=0.99\columnwidth]{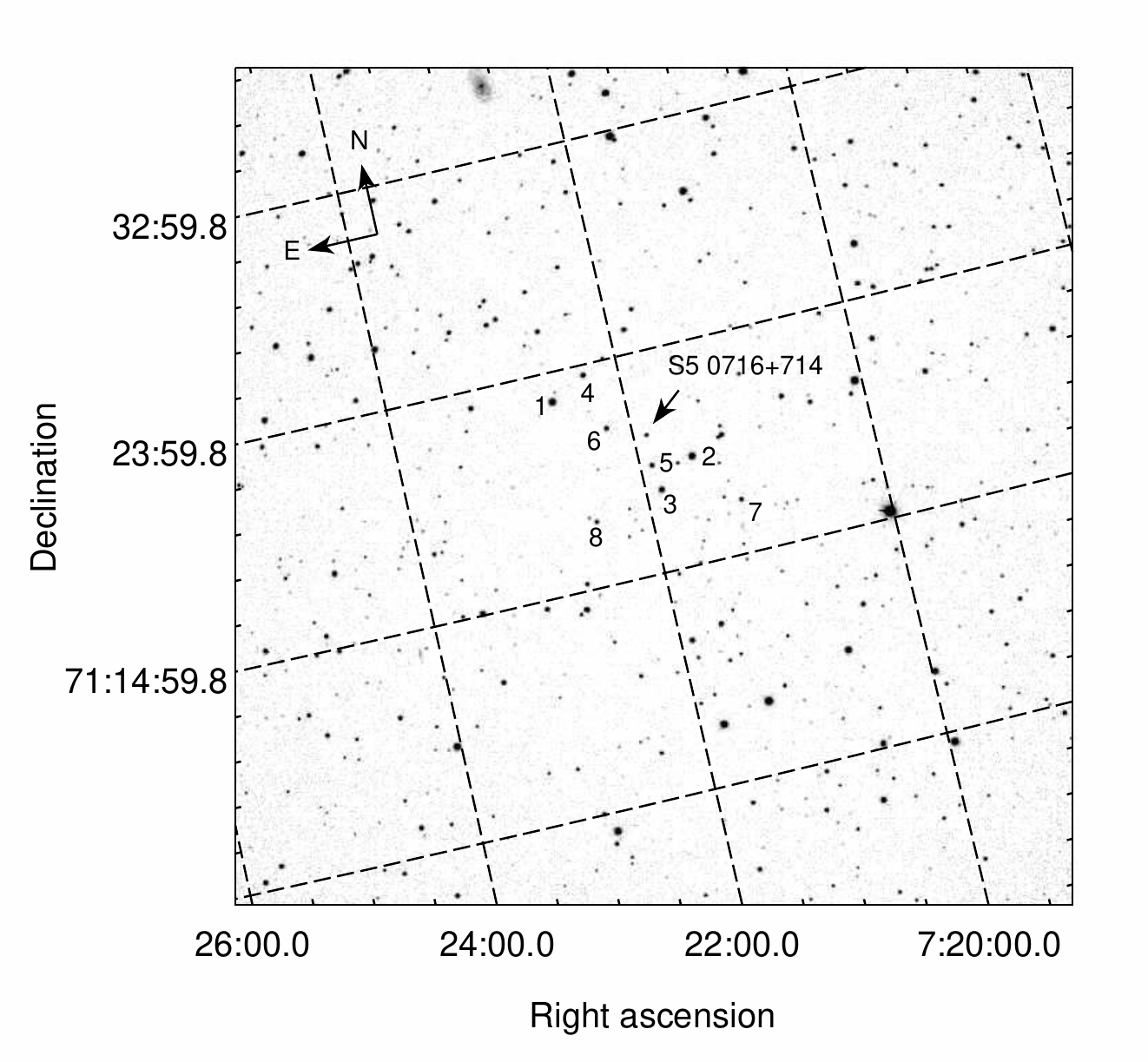}
\centering
 \caption{Finding chart of S5 0716+714 on the I band on the third night. The
 comparison stars are adopted from
 \citet{1998A&AS..130..305V}.\label{fig:finding chart}}
\end{figure}

\begin{figure}
\includegraphics[width=0.99\columnwidth]{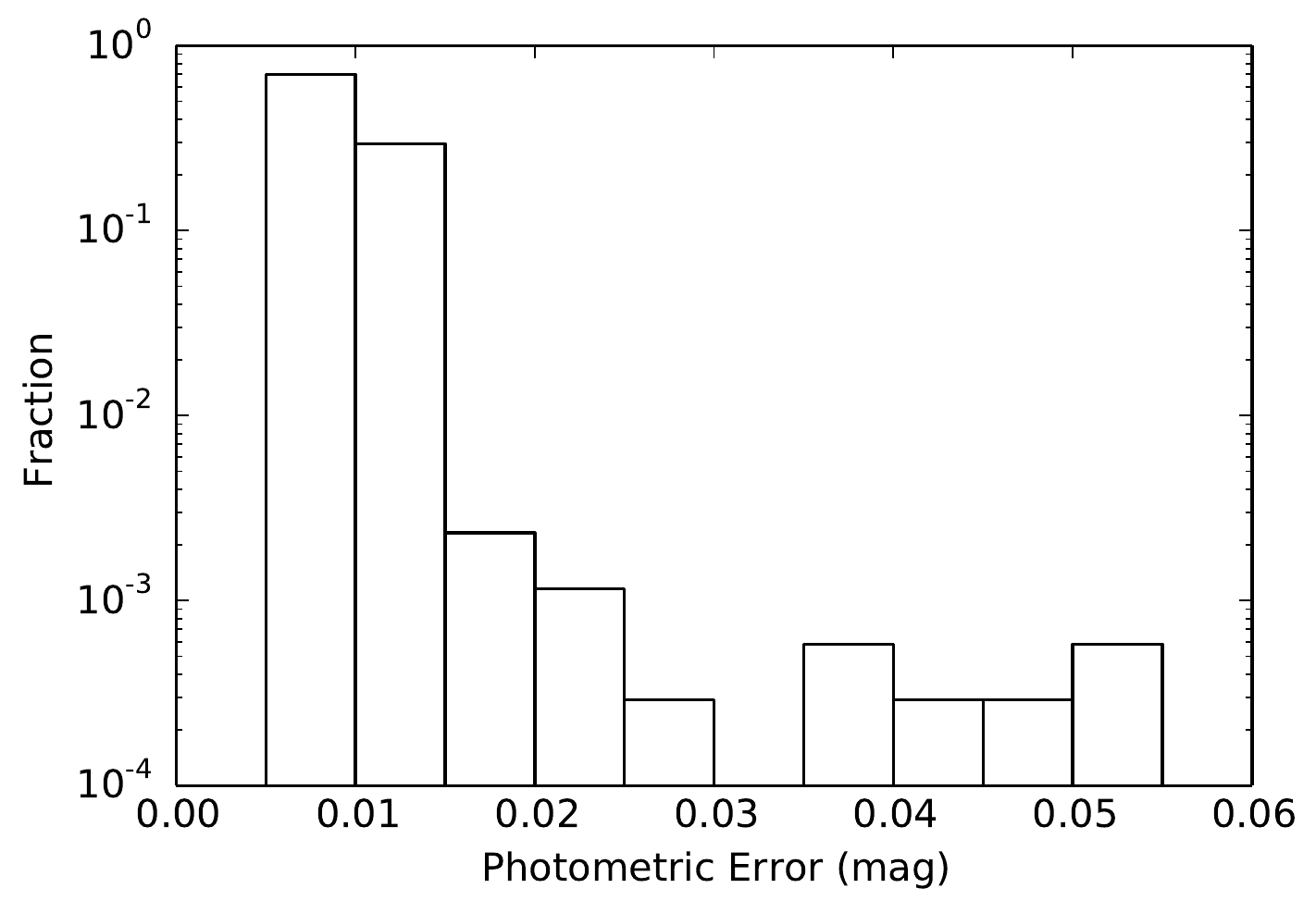}
\centering
 \caption{The distribution of photometry error of all observations. \label{fig:histogram}}
\end{figure}

\begin{figure*}
\includegraphics[width=0.99\textwidth]{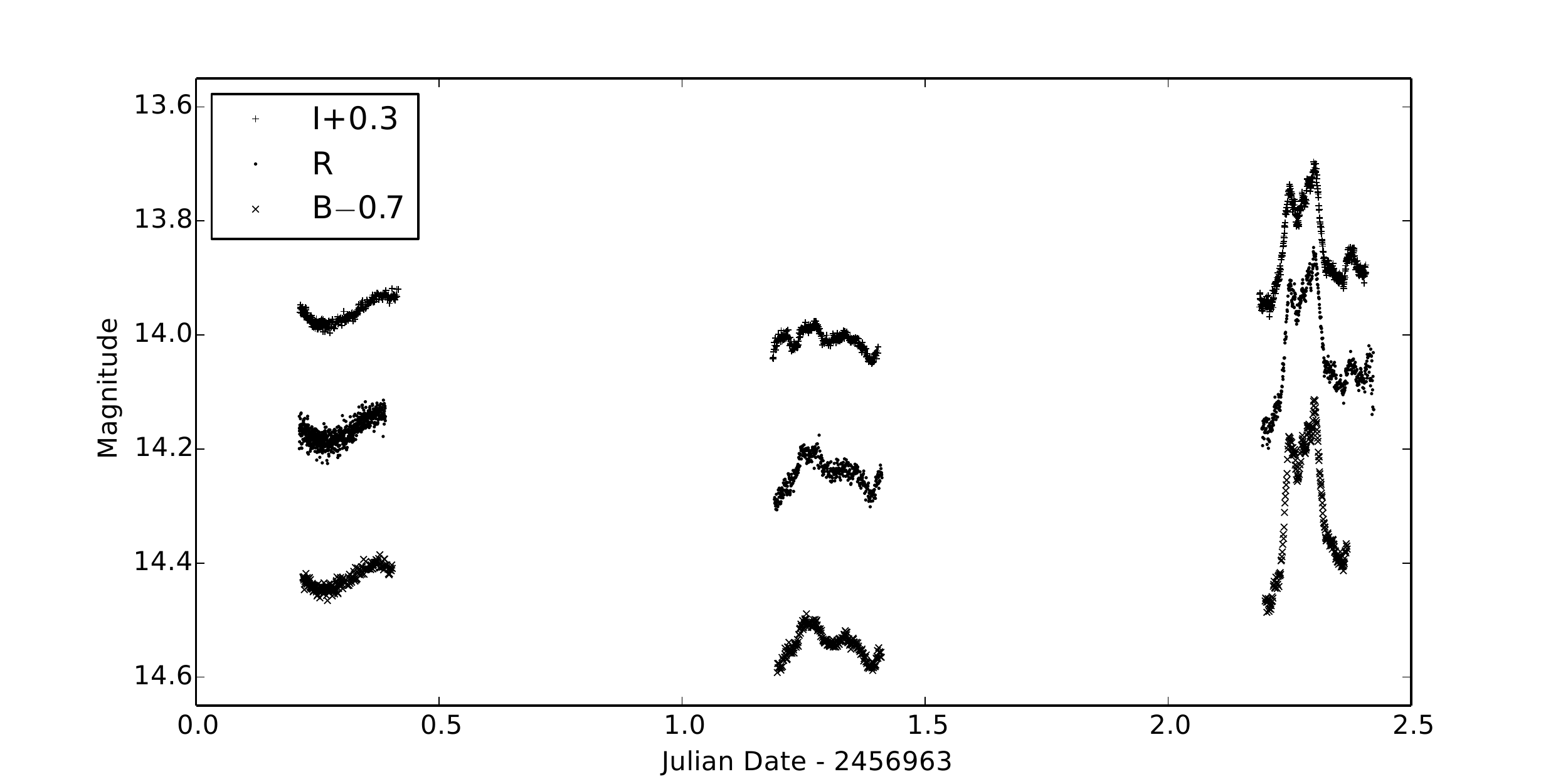}
\centering
 \caption{Overall light curves of S5 0716+714.\label{fig:overall}}
\end{figure*}

S5 0716+714 is one of the best studied BL Lac objects. Its redshift was
estimated as {$z$} = 0.31 $\pm$ 0.08 by \citet{2008A&A...487L..29N} with the photometric
detection of the host galaxy. More recently, a limit of 0.2315 {$< z <$} 0.322
was given by \citet{2013ApJ...764...57D}. Multi-wavelengths campaigns have been devoted to
study the SED of the source (e.g., \citealt{2006A&A...457..133F};
\citealt{2008A&A...487L..49G};
\citealt{2008A&A...481L..79V}). Because its peak frequency of the low-energy
component of SED lies between 10$^{14}$ and 10$^{15}$ Hz, the object was
classified as an IBL\citep{2008A&A...487L..49G}. 

Among all the most extremely variable blazars, BL Lac S5 0714+716 is one with
confirmed and documented variability at all wavelengths and on different
timescales ranging from minutes to decades. \citet{2013A&A...552A..11R}
investigated radio to gamma-ray variability of the source from 2007 to 2011,
and found correlation between radio and gamma-ray variations.
\citet{2012A&A...543A..78L} found its intra-day variability (IDV) and
long-term flux variations at 4.8 GHz. It is widely confirmed that the source
performs almost continuous micro-variability activity with a duty circle close
to 1 (e.g. \citealt{1996AJ....111.2187W}; \citealt{2010AAS...21642011W};
\citealt{2007AJ....133.1599W}; \citealt{2011ApJ...731..118C};
\citealt{2012AJ....143..108W};  \citealt{2012MNRAS.421.3111Z};
\citealt{2014MNRAS.443.2940H}).
In order to trace such very fast variations, high time resolution is required
for the monitoring campaigns. For example, \citet{2013A&A...558A..92B} conducted a 72-h WEBT
microvariability observation with exposure time from 30-150s. The exposure
time for \citet{2014MNRAS.443.2940H} is 15 to 300s, and the time for
\citet{2013ApJS..204...22D} is 30 to 480s. 

Attempts have been made to detect time delays in different frequency of S5
0716+714: time lags among radio, optical, X-ray and gamma-ray bands have been
calculated (e.g. \citealt{2013A&A...552A..11R}; \citealt{2014ApJ...783...83L}) as several
days. For optical bands, a possible lag of about 11 minutes was found by
\citet{2009ApJS..185..511P} between the {\it B} and {\it I} bands.
\citet{2010ApJ...713..180Z} claimed lags of
a few minutes between different optical bands. \citet{2012AJ....143..108W}
found variations in the {\it B'} and {\it V'} bands lead that of the {\it R'}
band by about 30 minutes on one night. However, the time delays are sometimes
too short to be detected. \citet{2012MNRAS.425.1357G} tried to calculate the
time delay between {\it V} and {\it R} bands, but the ZDCF peak is too close
to zero and the exposure time of their observation is 180s which may be longer
than the lag itself. The high temporal resolution of our observation can help
us obtain more convincing optical time delay of S5 0716+714.
 
The mechanism leading to blazar variability remains unsettled. One of the most
favored senarios is the shock-in-jet model: shock arises from the root of the
jet and propagate outwards, accelerating electrons and compressing magnetic
fields (\citealt{1985ApJ...298..114M}; \citealt{1991A&A...241...15Q}). However, other scenarios like
microlensing effects (\citealt{1986A&A...157..383N}), geometric effects
(\citealt{1992A&A...255...59C}) and
interstellar scintillation (\citealt{2001ApJ...550L..11R}; \citealt{2012A&A...543A..78L}),
etc. also contribute in variability. \citet{2014MNRAS.443.2940H} ascribed the IDV of S5
0716+714 to the intrinsic shock-in-jet models and geometric effects.
\citet{2012MNRAS.421.3111Z} argued that the turbulent process leads to
variability. In fact, the mechanism of variability seems complicated than just
one simple explanation.

In order to trace the intra-day optical variability of S5 0716+714 in multiple
color and search for the possible time delays, we carried out short term
observations in the {\it B}, {\it R} and {\it I} broad bands with three
telescopes on the nights of November 1$^{st}$, 2$^{nd}$ and 3$^{rd}$, 2014.
Our temporal resolution are 73s, 34s, 58s for three days average of the {\it B}, {\it R} and {\it I} bands,
which are higher than some of other campaigns. Hence, we can obtain a more
detailed record of the microvariability and search for possible optical
inter-band time delay. 
The observation, data reduction, and results are presented and discussed in
the following sections. 

\section{Observations and data reduction}
We performed a three-day observation of S5 0716+714 with three telescopes at
the Xinglong Station of the National Astronomical Observatories of China
(NAOC). The detail of the telescopes is shown in Table.\ref{table:telescope}. Observations
started from the midnights of November 1$^{st}$, 2$^{nd}$ and 3$^{rd}$, 2014
simultaneously on three telescopes, ended on the next morning before 7am, the
average time span of each night are 4.50 hr, 5.26 hr, 5.10 hr, respectively.
The exposure time were set as 70s, 10s, 40s for the filters {\it B}, {\it R}
and {\it I}, respectively, which were adjusted referring to the seeing, moon
phase, weather conditions then, but never exceed 80s.
The average exposure time for each band were 62.5s, 16.3s, 36.4s and the average temporal
resolution were 73s, 34s, 58s. We used the IRAF\footnote{IRAF is
distributed by the National Optical Astronomy Observatories,
which are operated by the Association of Universities for Research
in Astronomy, Inc., under cooperative agreement with the National
Science Foundation. (http://iraf.noao.edu)} to conduct the data reduction procedure. Four stars
(2, 3, 5, 6 in fig.\ref{fig:finding chart}) were selected as comparison stars. The standard
magnitudes of the stars in the {\it B}, {\it R} and {\it I} bands are given
by \citet{1998A&AS..130..305V} and \citet{1997A&A...327...61G} (listed in
Table.\ref{table:magnitude}).  The magnitude of S5 0716+714 was measured
relative to the four comparison stars. The aperture radii, inner radius of the
sky annuli, the width of the sky annuli are set according to the FWHM derived
from the images.

\begin{table}
\centering
 \begin{minipage}{140mm}
  \caption{Parameters of the telescopes\label{table:telescope}}
  \begin{tabular}{@{}lcrr@{}}
  \hline
   Telescopes  \footnote{All three telescopes are located at the Xinglong Station of the
   
    National Astronomical Observatories of China (NAOC).}      &  Filter & CCD resolution &  CCD view    
\\

 \hline
 60cm reflecting telescope &  {\it R} &$1^{''}.06$/pixel &$11^{'}\times11^{'}$\\
 80cm Cassegrain reflector & {\it I} &$0^{''}.51$/pixel &$18^{'}\times18^{'}$\\
 85cm Cassegrain reflector &{\it B} &$1^{''}$/pixel &$33^{'}\times33^{'}$\\
 \hline
\end{tabular}
\end{minipage}
\end{table}

 %We set the aperture radii as 1.5$\times$FWHM, the inner radius of sky annulus 5$\times$FWHM and the width of the sky annuli 2$\times$FWHM. The value of the FWHM and STDDEV (needed by IRAF) is derived from the images of separate night. 
  %Difference between the instrumental aperture magnitude and the standard magnitude of the four reference stars is calculated and averaged, then we derived the actual magnitude of the object from its instrumental aperture magnitude. 

 \begin{table}
\centering
 \begin{minipage}{140mm}
  \caption{The standard magnitudes of the four reference stars\label{table:magnitude}}
  \begin{tabular}{@{}cccc@{}}
  \hline
   Reference stars  \footnote{Standard magnitudes are given by
   \citet{1998A&AS..130..305V} and\\ \citet{1997A&A...327...61G}.}      &      {\it B} &{\it R} &{\it I}       
\\
 \hline
 2& 12.02 &11.12 &10.92 \\
 3 & 13.04&12.06&	11.79 \\
 5&14.15&13.18&12.85 \\
 6&14.24&13.26&	12.97 \\
 \hline
\end{tabular}
\end{minipage}
\end{table}

The photometric errors of our observations are all below 0.07 mag, most of
which are around 0.01 mag. The distribution of photometric error is shown in
fig.\ref{fig:histogram}. The
proportion of error over 0.02 mag is only 0.0026 representing 9 data points in {\it
R} band on the end of the second night. Their relatively large photometric
errors were the result of the sky background getting brighter at the beginning
of morning twilight. We obtained the
total number of the data points as 3444: 718 images for the {\it B} band, 1717
for the {\it R} band, and 1009 for the {\it I} band.

\section{Light Curves}

\begin{table*}
\centering
 %\begin{minipage}{160mm}
  \caption{Results of one-way ANOVA test and $\chi^2$ test of light curves in
  each band for each day\label{table:test result}} 
  \begin{tabular}{@{\extracolsep{18pt}}cccrccccc@{}}
  \hline
  JD&Band&$N$&\multicolumn{2}{c}{One-way ANOVA}  & \multicolumn{2}{c}{$\chi^2$
  test}&Result\\
  %\footnote{Reference: TESTING TESTS ON ACTIVE GALACTIC NUCLEI MICROVARIABILITY  de Diego AJ. 2010}\\
  \cline{4-7}
  & & & $F_{\nu_1,\nu_2}$& $F_{\nu_1,\nu_2}^{0.001}$& $\chi^2$&$\chi^2_{0.001,\nu}$ &\\
 \hline
 2456963&B&198& 91.6143&3.0172&233.6034&264.0754&V$^{*}$\\
 2456963&R&797&182.3321&3.1388&870.5844&925.0195&V$^{*}$\\
 2456963&I&259&378.7852&3.0904&776.8546&333.9289&V \\
 2456964&B&234&148.3069&2.9881&426.4335&305.4426&V \\
 2456964&R&434&142.3280&3.0352&720.6714&529.6648&V \\
 2456964&I&275& 90.9400&3.0822&628.1363&352.0720&V \\
 2456965&B&233&321.4710&3.2459&7455.161&304.2994&V \\
 2456965&R&491&401.1203&3.0261&8631.511&592.4640&V \\
 2456965&I&475&360.2808&3.0285&11665.65&574.8719&V \\
 \hline
\end{tabular}
%\end{minipage}
\end{table*}

\subsection{Variability Test}
In order to verify the IDV of S5 0716+714, we performed two robust
statistic methods introduced by \citet{2010AJ....139.1269D}: one-way ANOVA test and $\chi^2$ test.
In one-way ANOVA test, we sampled observation data every 30 minutes, so the
total number of groups $k\sim11$ in each band every night. In $\chi^2$ test,
we considered a correction factor $\eta=1.5$ \citep{2004MNRAS.350..175S,2008AJ....136.2359G},
and we assumed the rescaled error $\sigma_{\mathrm{real}}=\eta\sigma$ as
real error in test calculation.

The results of both tests are presented in Table.\ref{table:test result}. All 9 $F$ values of one-way
ANOVA test are significantly larger than critical values of $\alpha=0.001$. However, 2 of
9 $\chi^2$ values are slightly smaller than critical values of $\alpha=0.001$,
which were noted as V$^{*}$ in the column of result. Those with both F value
and $\chi^2$ value over critical values were noted as V (variable).
This incompatibility can be ascribed to the discrepency of power in these
methods and was discussed by \citet{2010AJ....139.1269D}. As a result, 
we can conclude that IDV can be found on all three nights in all three bands.

\subsection{Intra-day Variability}
The overall light curves in the {\it B},  {\it R} and {\it I} bands are
presented in fig.\ref{fig:overall}. Magnitudes of the {\it B} and {\it I} bands are shifted
closer to one another in order to provide a better view. 
On the first night, the time interval between the maximum and the minimum of
the variation is about 2.6 hr in the {\it I} band, during which the brightness
varied by 0.08 mag. The pattern of the curve seems approximately a sinusoidal
curve, such a curve of this source was also found by
\citet{2005AJ....129.1818W}.
On the second night, two small flares can be seen in the light curves
in the {\it R} and {\it B} bands, but it seems there is an extra flare before
the two flares in the {\it I} band. The cause of such difference is unclear,
but at least it should not be attributed to our data reduction. Microlensing
effect and interstellar scintillation should be excluded since the former
influences emission in all wavelengths \citep{1995ARA&A..33..163W} while the latter would not affect the optical variation.  

Light curves on the third night reveals violent variability in all three bands
compared with earlier days (see fig.\ref{fig:overall}). A double-peak structure can be seen
clearly with five small sub-peaks overlapped on. At first, the object turned
bright dramatically to the first peak, then it came down a little before
reaching another peak again. The second peak is brighter than the first, and
it is only 1.3 hr between the two peaks. The brightness turned down quickly
followed by a small flare at the end of the observation. By visual inspection,
we can see quasi-periodic oscillations e.g. three flares on the second peak
separated by about 20 minutes.

\begin{figure}
\includegraphics[width=0.99\columnwidth]{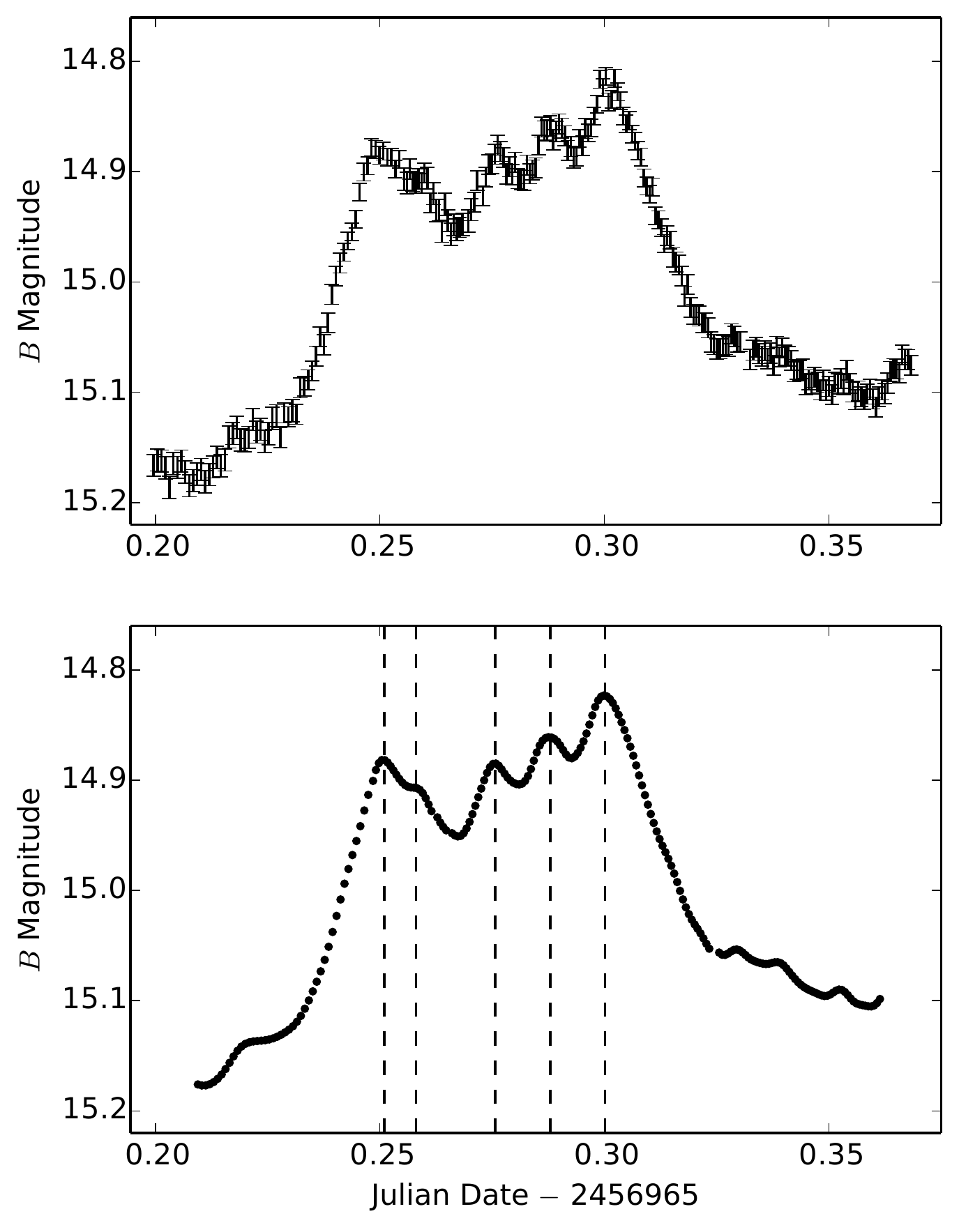}
\centering
 \caption{Upper panel: Light curve of S5 0716+714 on the third night in {\it
 B} band. A maximum changing rate of 0.347 mag h$^{-1}$ ({\it B} band) was
 detected.
 Lower panel: Light curve after smoothed by FIR filter, five flares are marked
 by vertical dashed lines, and demonstrate QPOs.\label{fig:day3}}
\end{figure}

To expose these peaks more precisely, we filtered the original light curve with
a finite impulse response (FIR) filter.
The length of filter and cutoff frequency were set to be 20 and 0.1
min$^{-1}$, separately. The consequence was shown in lower panel of
fig.\ref{fig:day3}. Five
flares was marked by vertical dashed lines. The time intervals between each
flare are 10.14, 25.47, 17.62 and 17.62 minutes, respectively. Despite the second
flare shifted to an earlier time, the separation between
first and last flare is $\sim71$ minutes, which is almost 4 times of the last
two intervals. This 17.6 minutes period is close to $\sim25$ minutes
\citep{2009ApJ...690..216G}and $\sim15$ minutes \citep{2010ApJ...719L.153R}
in the historical observation of QPO of this source.

Specifically, the IDV of the third day is remarkable for its large amplitude
and fast change rate which exceeds many of its historical records. In
literature, S5 0716+714 was observed to vary by 0.117 mag in 1.1 hr in {\it c}
band \citep{2013ApJS..204...22D}. \citet{2002PASA...19..143N} found a maximum rising rate of 0.16 mag
h$^{-1}$. The fastest changing rate of almost 0.38 mag h$^{-1}$ was detected
by \citet{2011ApJ...731..118C}. Our observation reveals a maximum changing rate of 0.347 mag
h$^{-1}$ ({\it B} band) which has been quite a high rate so far. In other
blazars, the change rate of variation could be even higher i.e. PKS 2155-304
was detected to have a very fast variability rate of 0.43 mag
h$^{-1}$\citep{2014A&A...562A..79S}. 

In general, the trends of light curves in different bands are basically the
same but with different amplitude. On the fist two days, amplitudes of light
curves in the {\it B} and {\it R} bands are slightly larger than that of the
{\it I} band. The amplitude of magnitude on the third day decreases with
wavelengths: the rising amplitude is 0.27 mag for {\it B} band but 0.25 mag
for {\it I} band. This trend has been confirm by a number of authors for
different blazars (e.g. \citealt{2012AJ....143..108W};
\citealt{2012MNRAS.425.1357G};
\citealt{2013ApJS..204...22D}; \citealt{2014AJ....148..110M}). The trend, however, may lead to the special color behavior as discussed in the next section.

\section{Color behavior and spectral hysteresis}
\begin{figure*}
\includegraphics[width=0.68\columnwidth]{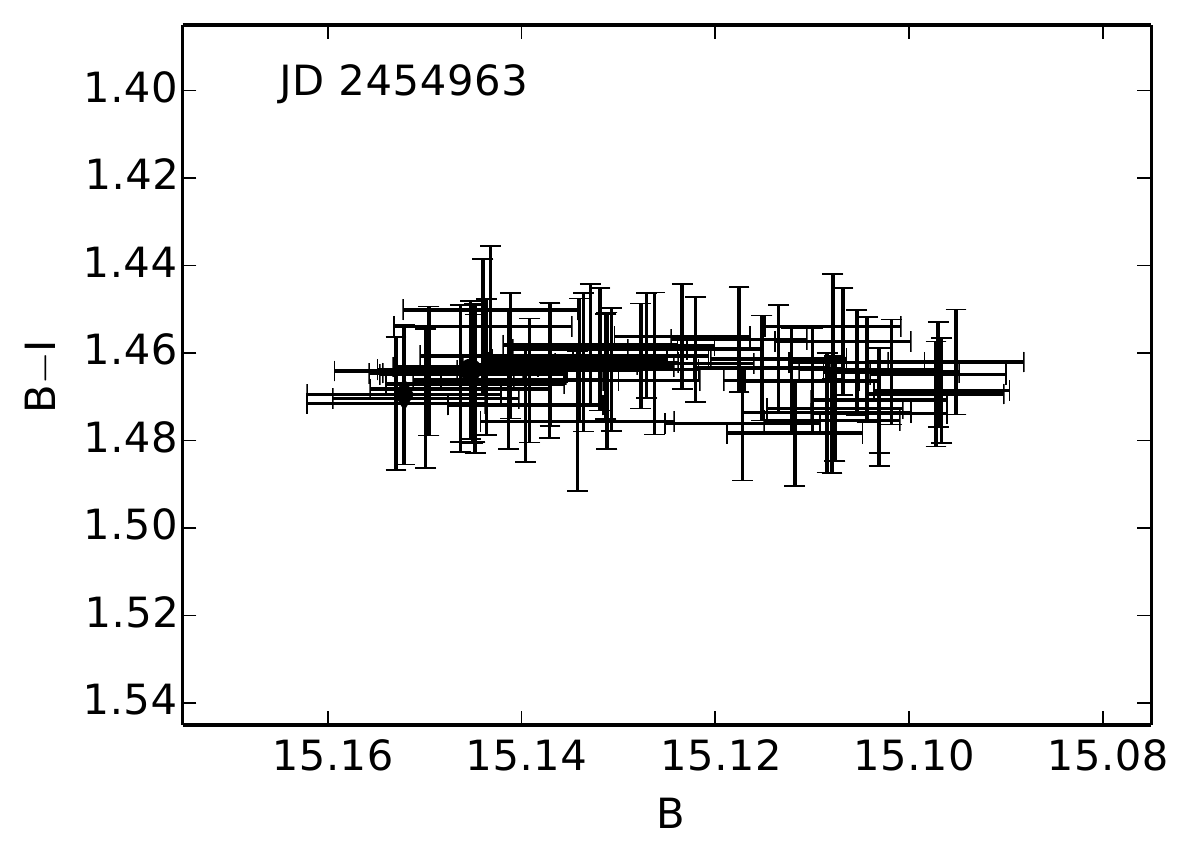}
\centering                                        
\includegraphics[width=0.68\columnwidth]{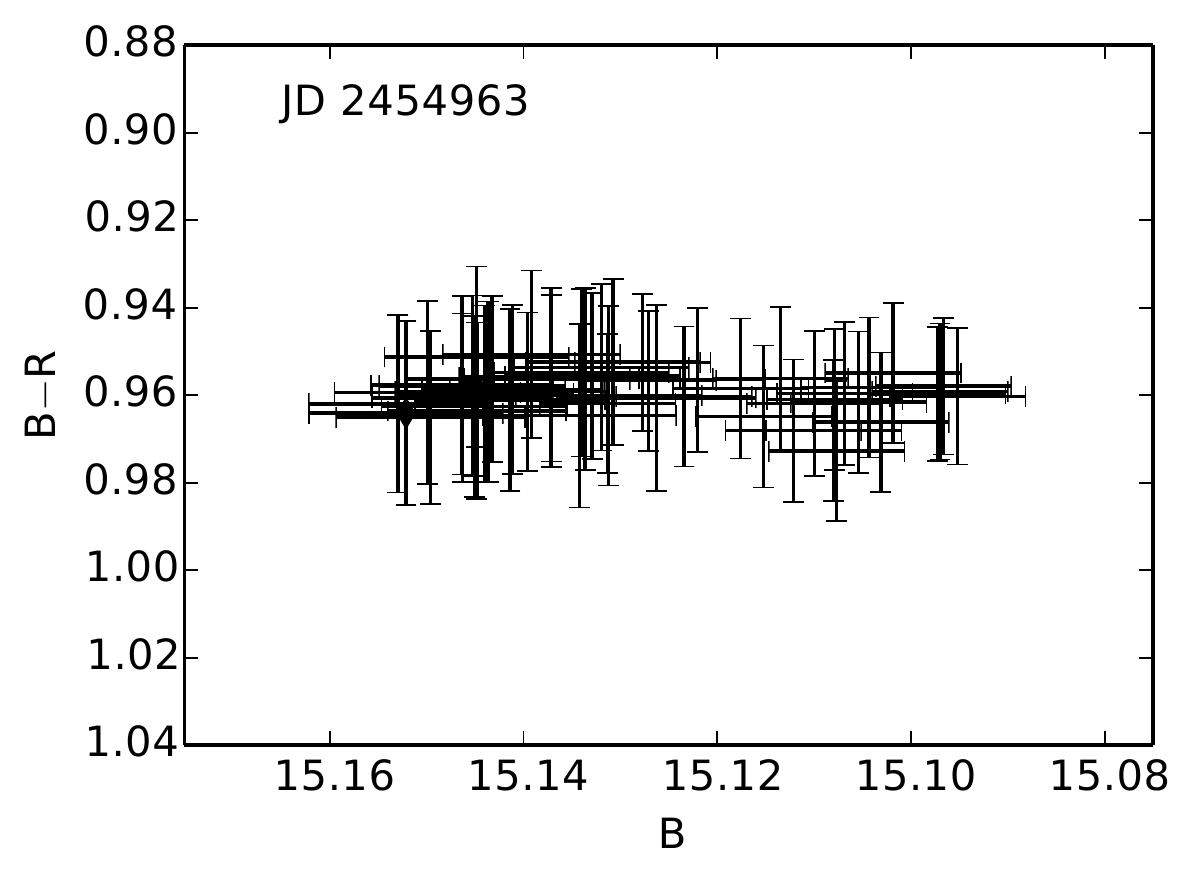}
\centering                                        
\includegraphics[width=0.68\columnwidth]{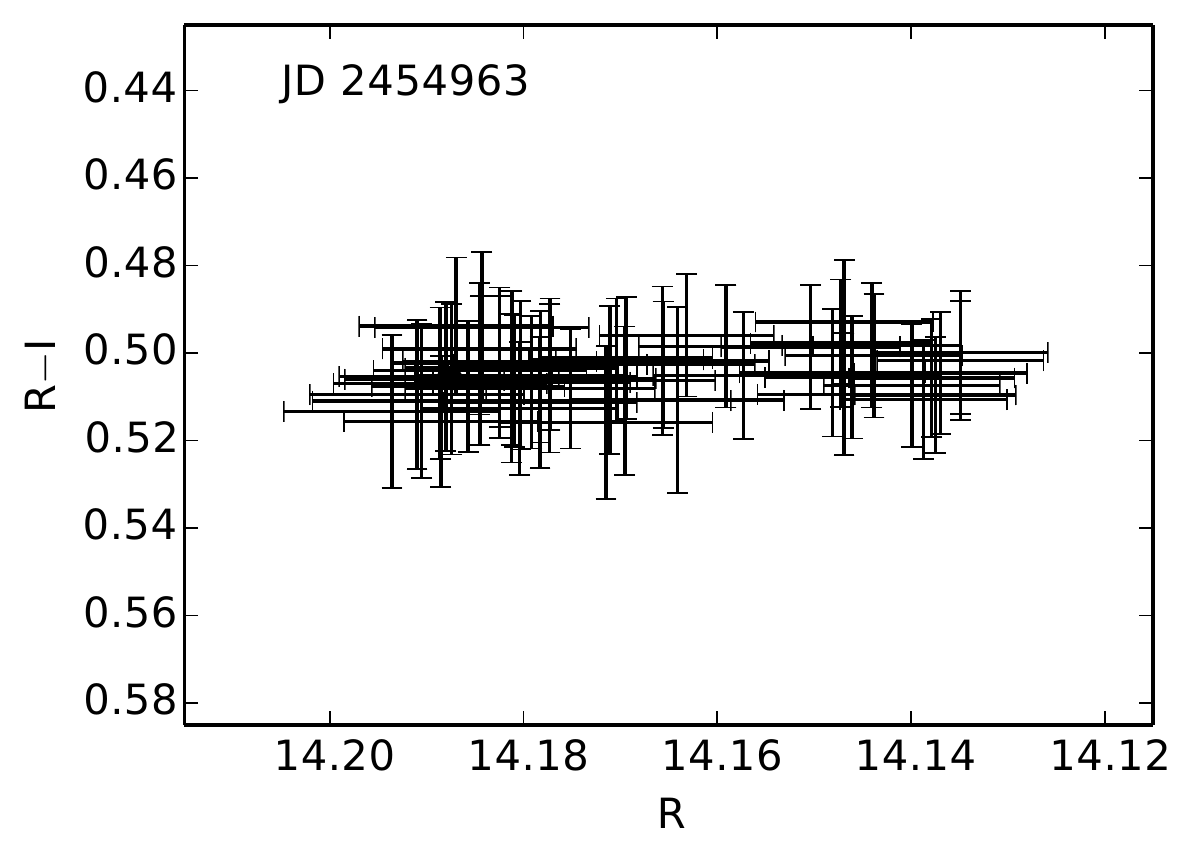}
\centering
\includegraphics[width=0.68\columnwidth]{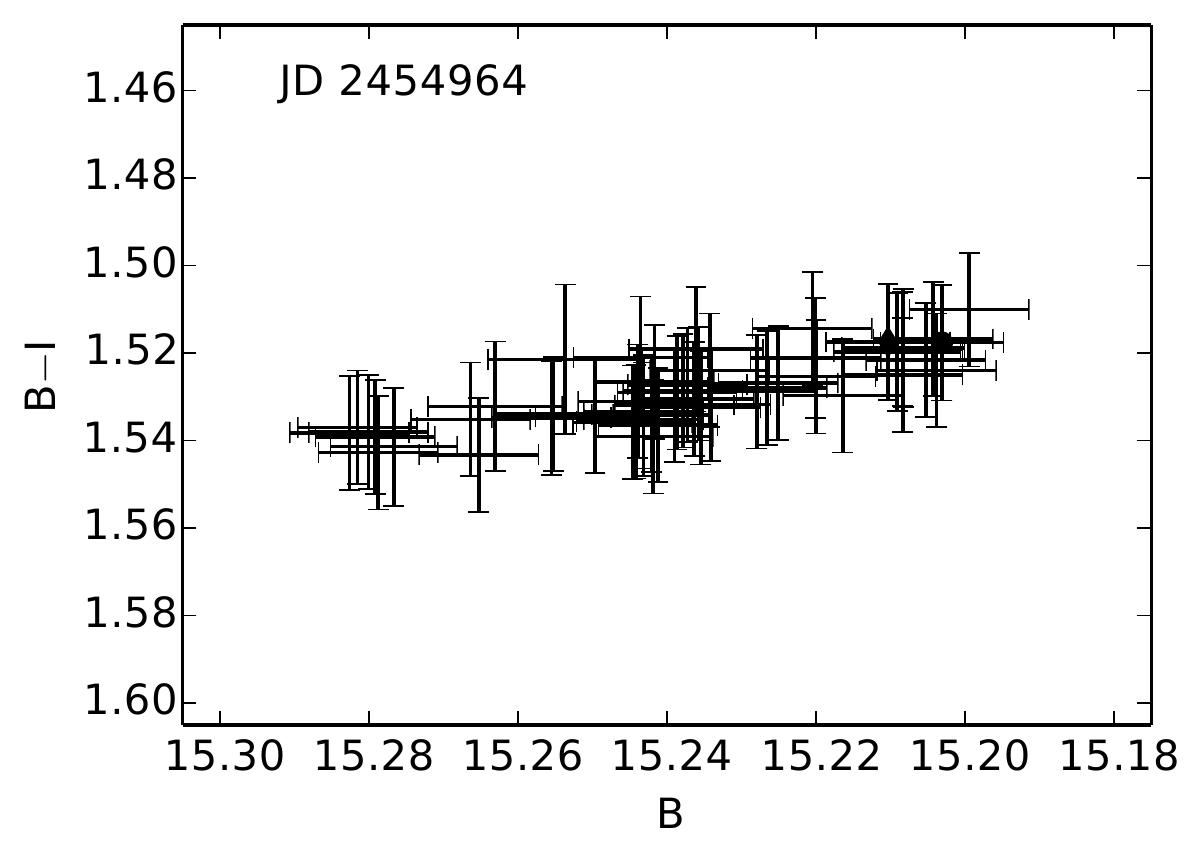}
\centering
\includegraphics[width=0.68\columnwidth]{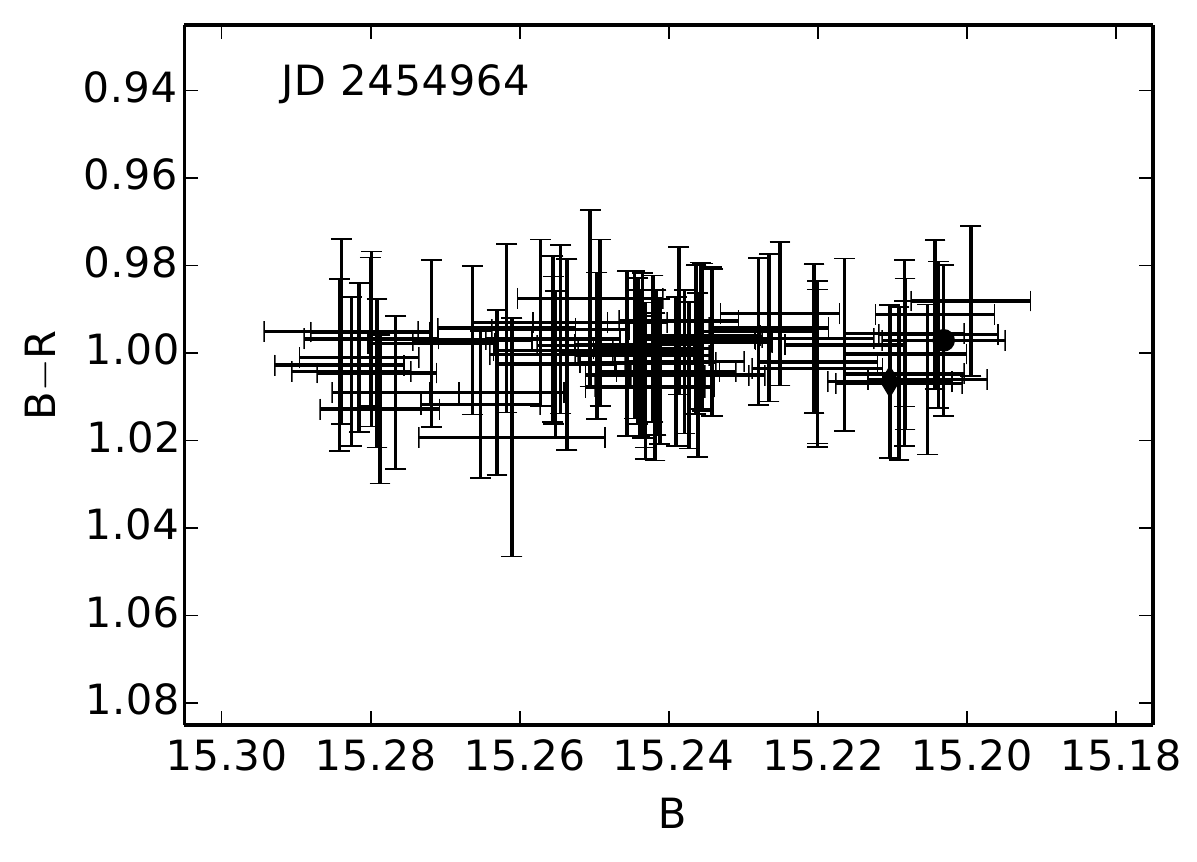}
\centering
\includegraphics[width=0.68\columnwidth]{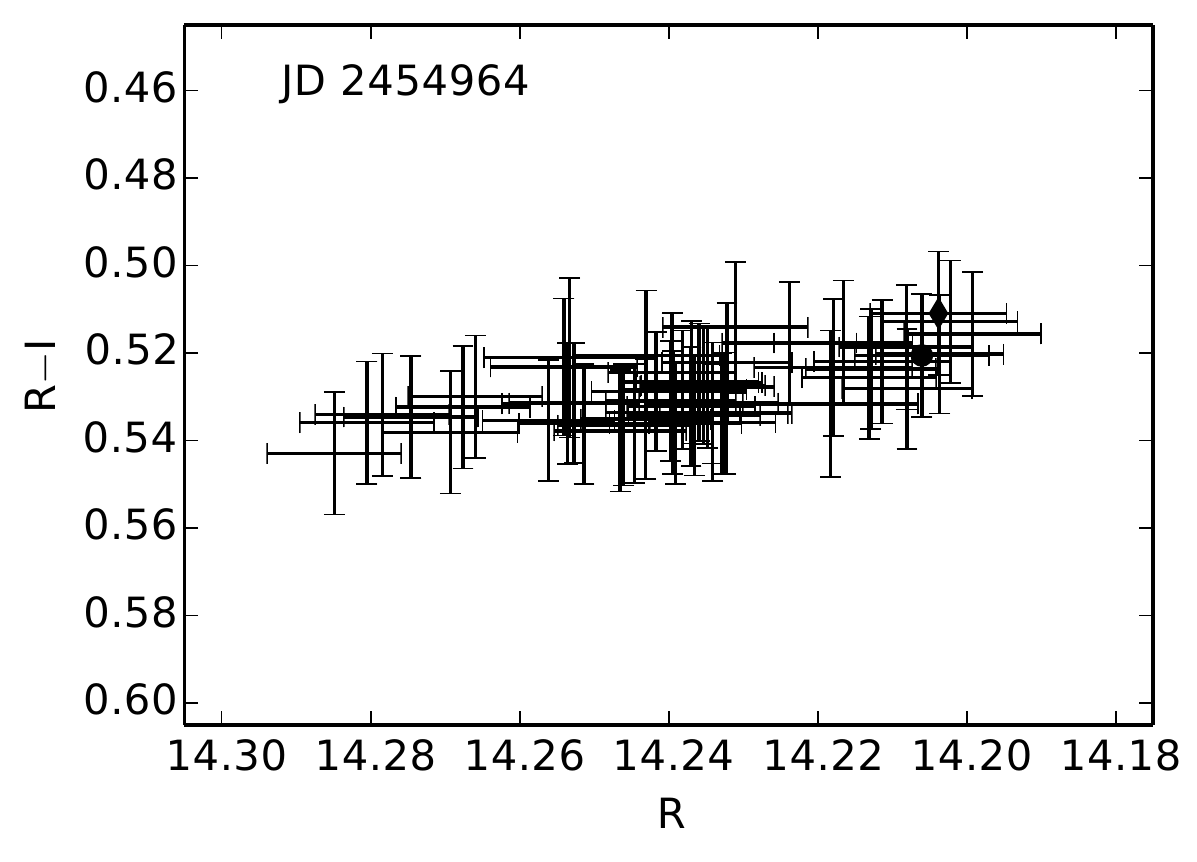}
\centering
\includegraphics[width=0.68\columnwidth]{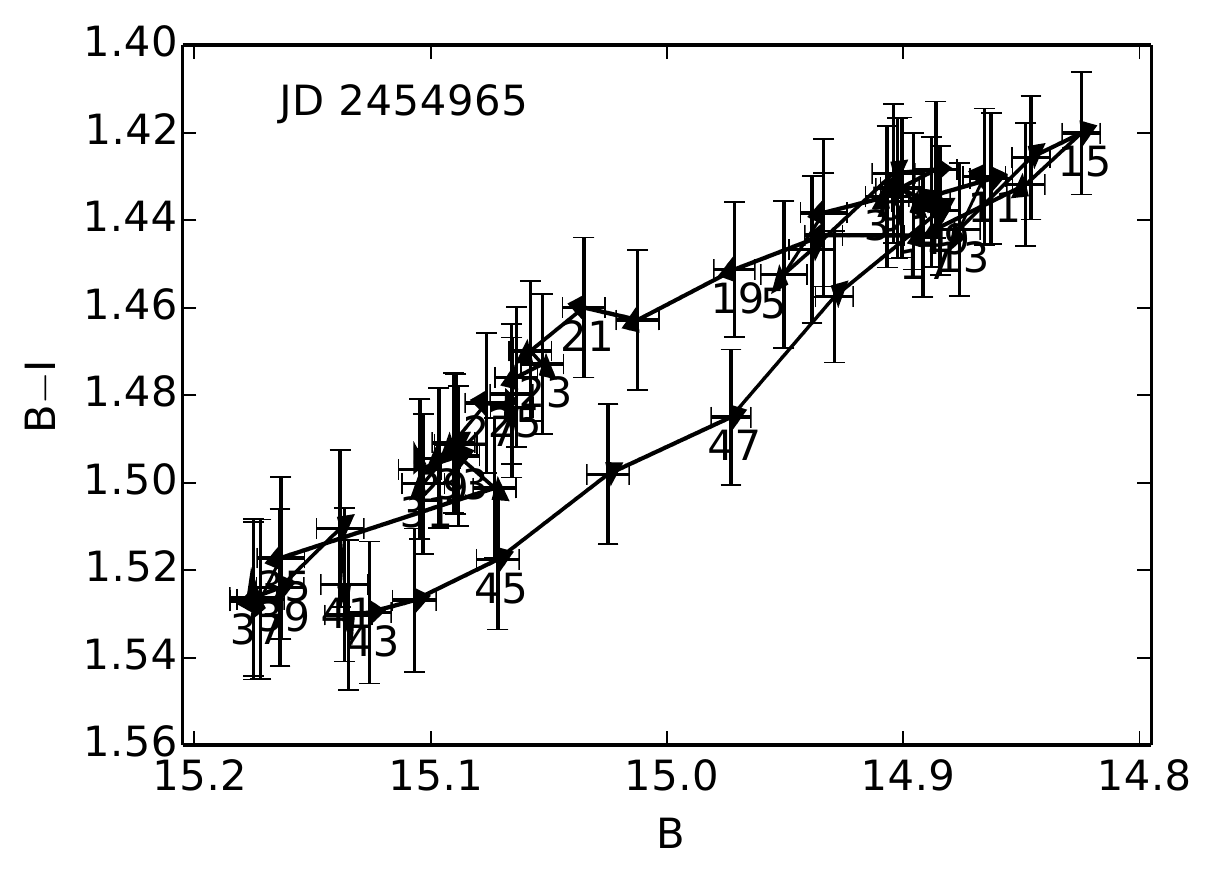}
\centering                 
\includegraphics[width=0.68\columnwidth]{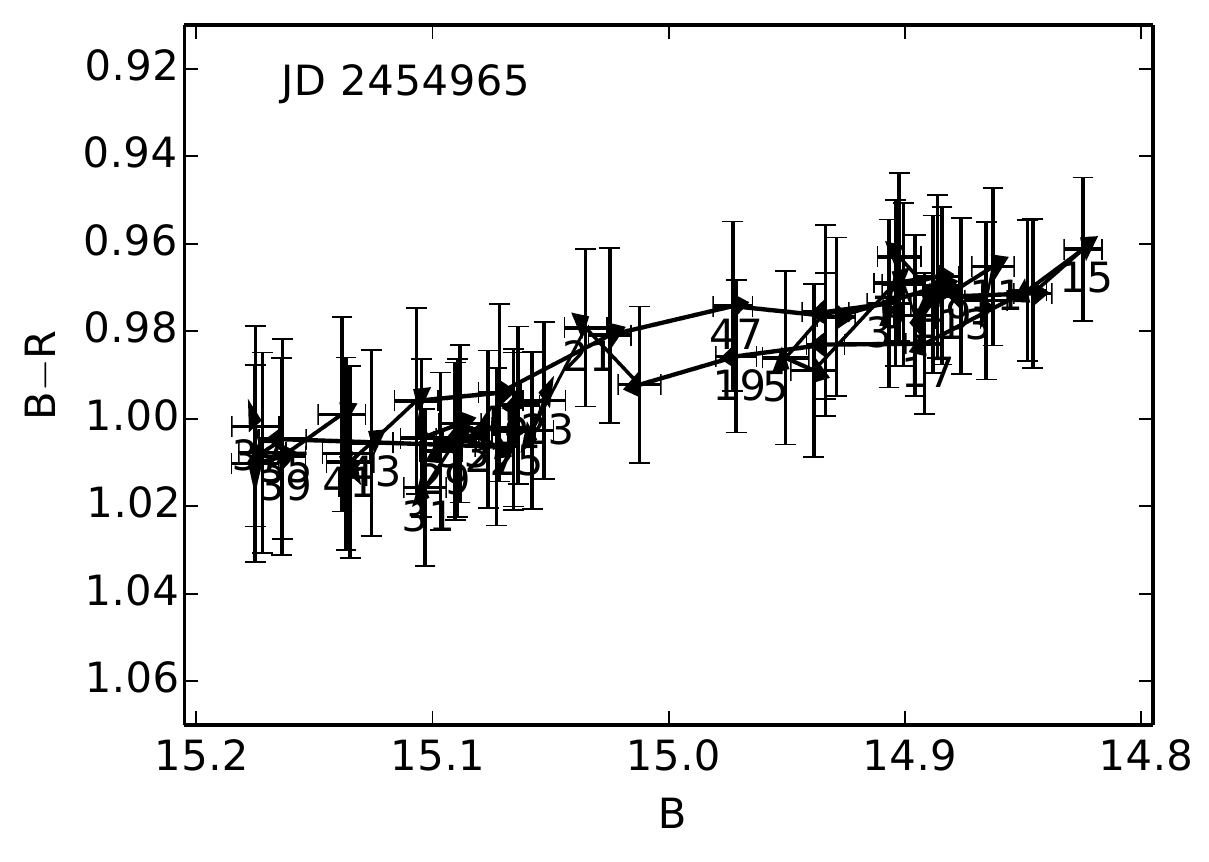}
\centering                 
\includegraphics[width=0.68\columnwidth]{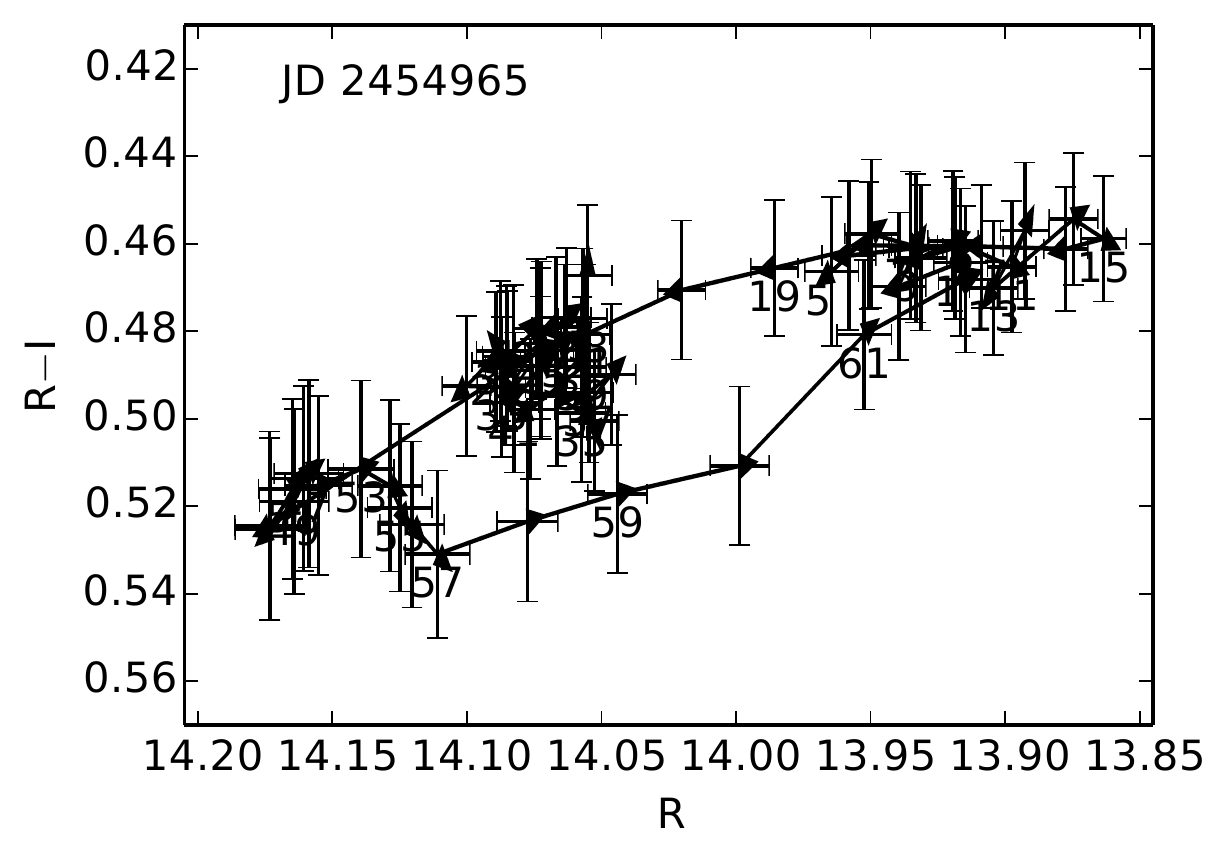}
\centering
 \caption{Color behaviour of S5 0716+714 on three nights. Top three diagrams
 demonstrate the absence of color variation on the first night; three in the
 middle reveal BWB on the second night; color evolution on the third night is
 presented on the bottom three diagrams as clear loop paths, arrows are
 plotted to show the direction of the loop. \label{fig:color}}
\end{figure*}

We investigated the color-magnitude relationship for each separate night.
Specificly, we use {\it B}$-${\it R}, {\it R}$-${\it I} and {\it B}$-${\it I}
to represent the color of the target, trying to show how they are  related to
{\it B}, {\it R} and {\it I} magnitudes, respectively. The results displayed
in fig.\ref{fig:color} reveal that the color of the object remained almost unchanged during
the first night, which may explain the sinusoidal curve on this night: this
has been reported occasionally before (e.g. \citealt{2005AJ....129.1818W};
\citealt{2009ApJS..185..511P}). Such achromatic variation is probably caused
by the lighthouse effect as a geometric factor in shock-in-jet model according
to \citet{1992A&A...255...59C}: when enhanced particles are moving relativistically toward an observer on helical trajectories in the jet, flares will be produced by the sweeping beam whose direction varies with time.  

On the second night, a slight bluer-when-brighter (BWB) chromatism can be
seen; on the third night, however, the object exhibited strong BWB chromatism.
Such BWB trend is common for many blazars, especially for BL Lacs. S5 0716+714
has shown strong BWB in both intra-night and inter-night timescales in
literature (\citealt{2003A&A...402..151R}; \citealt{2007AJ....133.1599W};
\citealt{2011ApJ...731..118C}; \citealt{2013ApJS..204...22D};
\citealt{2014MNRAS.443.2940H}). The mechanism of BWB was simulated and
attributed to the difference of the amplitudes and cadences at different
wavelengths \citep{2011AJ....141...65D}. For FSRQs, both BWB and
redder-when-brighter trend can be found, and it is hard to conclude which
dominates (\citealt{2011A&A...534A..59G}; \citealt{2006A&A...450...39G};
\citealt{2006MNRAS.371.1243H}).

In fig.\ref{fig:color}, a loop-like pattern can be seen on the third day's color-magnitude
diagrams of {\it B}$-${\it I} vs {\it I} and {\it R}$-${\it I} vs {\it I}. By
examining the order of the points, we found the loop in the third day's
color-magnitude diagram traced anti-clockwise. It is the first time that the
intra-day spectral hysteresis loop has been found so obviously in the optical
band. Such loop is not a general feature of color evolution.
\citet{2012ApJ...756...13B} claimed OJ 287 move around on the circular locus
in color-magnitude space. \citet{2013ApJS..204...22D} found an anti-clockwise
loop in intra-day color behavior of S5 0716+714. \citet{2006A&A...448..143X}
have reported the similar pattern in optical band. In other regime like
gamma-ray and X-ray, the loop path has been reported frequently (e.g.
\citealt{1993ApJ...404..112S};
\citealt{1996ApJ...470L..89T};\citealt{2000MNRAS.312..123M};
\citealt{2007ApJ...657..760A}; \citealt{2010ApJ...715..554K}; 
\citealt{1999ApJ...527..719Z}; \citealt{2002ApJ...572..762Z}). 

\section{Cross-correlation analysis and time lags}
We performed a cross-correlation analysis to search for the possible lag
between variations in different bands. There are some cross-correlation
methods like the discrete correlation function (DCF) method
\citep{1992A&A...255...59C} and the z$-$transformed discrete correlation
function (ZDCF) method \citep{1997MNRAS.284..967A}. Here we use the
interpolated cross-correlation function (ICCF) method to estimate the lags and
their errors \citep{1987ApJS...65....1G}. This is a flux-randomization (FR)
approach, the model-independent Monte Carlo realization was used to estimate
the error, and each realization is based on a randomly chosen subset of the
original data points, namely the random subset selection (RSS). Considering
the interval between our observations are small, the low-order interpolation
in ICCF can do a reasonable approximation to the true behavior of the light
curve (\citealt{1998PASP..110..660P}, \citealt{2004ApJ...613..682P}).

 \begin{table}
\centering
 \begin{minipage}{140mm}

  \caption{Time lags given by ICCF method (Unit: minutes)\label{table:lag}}   
  
  \begin{tabular}{@{}cccc@{}}
  \hline
    Date  &{\it B}-{\it R}  & {\it R}-{\it I}  &  {\it B}-{\it I}    \\
 \hline

Nov. 1$^{st}$&$-$3.538$\pm$5.973&$-$5.703$\pm$6.988&$-$9.165$\pm$8.271\\
Nov. 2$^{nd}$&\ 0.997$\pm$1.911&$-$0.429$\pm$3.002&$-$0.313$\pm$2.943\\
Nov. 3$^{rd}$&\ 0.336$\pm$0.500&\ 1.308$\pm$0.603&\ 1.445$\pm$0.511\\

 \hline

\end{tabular}
\end{minipage}
\end{table}

In our work, 5000 independent Monte Carlo realizations were performed on the
light curves of each day. The results are listed in the Table.\ref{table:lag}. As can be
seen in the table, lags of the first two days are with large relative error
(greater than 1), so the results are not quite reliable. However, on the third
night, especially for the {\it B} and {\it I}, {\it R} and {\it I}, the
relative errors become less than 50$\%$, which suggests variations in the {\it
R} and {\it B} bands are lagging behind that of {\it I} band by
(1.308$\pm$0.603) min and (1.445$\pm$0.511) min, respectively. Noting that the
temporal resolution of our observation is shorter than the lag, and the order
of variations in different bands is the same as suggested by its spectral
evolution in previous chapter, the time lags given by ICCF between the {\it B}
and {\it I}, {\it R} and {\it I} bands are thereby relatively convincing on
the third night.

So far, there has not been an exactly justifiable theoretical interpretation
of such spectral hysteresis. Among many attempts the one given by
\citet{1999APh....11...45K}
is mostly favoured: in the blazar emission models like the synchrotron
self-compton model \citep{1997A&A...320...19M}, spectral shape is influenced by particle
acceleration ($\tau_{acc}$), synchrotron cooling ($\tau_{syn}$) and intrinsic
variability ($\tau_{var}$, which determines the flare duration $t_{flare}$)
(\citealt{1998A&A...333..452K}; \citealt{1999APh....11...45K};
\citealt{2004ApJ...605..662C}; \citealt{2008ApJ...680L...9S}). The interaction
among different processes will possibly lead to spectral evolution hence a
loop-like path in a color-magnitude (or spectral-index-flux) diagram. A
clockwise path may be seen when $\tau_{syn}\ll\tau_{var}\ll\tau_{acc}$
\citep{1999APh....11...45K} e.g., OJ 287 \citep{2012ApJ...756...13B} and Mkn 421
\citep{1996ApJ...470L..89T}. As for
anti-clockwise paths as found in our observation and PKS
2155$-$304\citep{1993ApJ...404..112S}, the case seems to be
$\tau_{var}\approx\tau_{syn}\approx\tau_{acc}$, indicating changes propagate
from low energies to high energies (\citealt{2010ApJ...715..554K};
\citealt{2005A&A...431..391V}). Consequently, flares in higher energies may
lag behind that of the lower energies, which is in accord with both the
spectral hysteresis we found and the time lags we calculated with different
light curves. However, we still cannot draw any firm conclusion from a
hysteresis loop on the basis of the model above. \citet{2000ApJ...536..729L}
argued that the hysteresis loop is quite sensitive to various parameters in
Kirk's model, such as the overall injection energy \citep{2004ApJ...605..662C}.

\section{conclusions}
Our monitoring session targeted the BL Lac object S5 0716+714 was performed in
the {\it B}, {\it R} and {\it I} bands on November 1$^{st}$, 2$^{nd}$ and
3$^{rd}$, 2014 on three telescopes located at the Xinglong, China. The average
temporal resolution of our observation are 73s, 34s, 58s for the {\it B}, {\it
R} and {\it I} bands, respectively. They are set high to trace the intra-day
variation of the object and to search for the optical inter-band time delays.
IDV was found in each band of each day. Particularly, on November 1$^{st}$, we
found an achromatic sine curve suggesting a possible lighthouse effect in the
shock-in-jet model  \citep{1992A&A...255...59C}. On November 2$^{nd}$, light curve in the {\it
I} band shows an extra flare than the {\it B} and {\it R} bands. The case is
rare and the cause is unclear. On November 3$^{rd}$, we caught flares with
a very fast changing rate up to 0.347 mag h$^{-1}$ and a double-peak structure
with five $\sim17.6$ - period QPOs overlapped on.
This embodies the intense variability of the object on short timescale. 

We investigated the color behavior of S5 0716+714. The first night showed
marginal color variation, while the second night displayed BWB and then this
trend became stronger on the third night. Particularly, loop-like paths can be seen on the
{\it B}$-${\it I} vs {\it I}, {\it R}$-${\it I} vs {\it I} graphs. The loop
traced anti-clockwise, possibly implying occurrence of a flare propagates from
lower to higher energies, as particles are gradually accelerated into the
radiating window \citep{1998A&A...333..452K}. It also suggests variations in
the {\it B} band lagged behind that of the {\it I} and {\it R} bands. Then we
use ICCF method to statistically give an estimate of the values of time delay.
Time delays on the third night are (1.308$\pm$0.603) min for {\it R}$-${\it
I}, (1.445$\pm$0.511) min for the {\it B}$-${\it I}.  Considering the lags are
longer than the temporal resolution, the results on the third day are rather
plausible. 

Unfortunately, we did not find any monitoring project on S5 0716+714 with high
time resolution at high energies during our campaign, otherwise we could
further correlate our data with data in X-ray or Gamma-ray. 
High temporal resolution multi-wavelenth observation campaign may help
gain a much more comprehensive understanding of radiative
process and physical model of blazars.

\section{acknowledgement}
We thank the anonymous referee for insightful comments and suggestions that
helped to improve this paper.
Our work has been supported by the National Basic Research Program of China
973 Program 2013CB834900; Chinese National Natural Science Foundation grants
11273006, 11173016, and 11073023; and the fundamental research funds for the
central universities and Beijing Normal University.

\bibliography{draft0716}
\bibliographystyle{mnras}
\bsp    % typesetting comment
\label{lastpage}

\end{document}